\documentclass[aps,prc,twocolumn,amsmath,amssymb,superscriptaddress,showpacs,showkeys,preprintnumbers]{revtex4-2}
\usepackage{dcolumn}
\usepackage{graphicx,color}
\usepackage{multirow}
\usepackage{textcomp}
\usepackage{physics}
\usepackage{dsfont}
\usepackage{float}
\usepackage{comment}

\usepackage{relsize}
\usepackage{ragged2e}
\usepackage{subcaption}
\usepackage{adjustbox}
\usepackage[table]{xcolor}

\usepackage{tikz}	
\begin{document}
	\newcommand {\nc} {\newcommand}
	\nc {\beq} {\begin{eqnarray}}
	\nc {\eeq} {\nonumber \end{eqnarray}}
	\nc {\eeqn}[1] {\label {#1} \end{eqnarray}}
\nc {\eol} {\nonumber \\}
\nc {\eoln}[1] {\label {#1} \\}
\nc {\ve} [1] {\mbox{\boldmath $#1$}}
\nc {\ves} [1] {\mbox{\boldmath ${\scriptstyle #1}$}}
\nc {\mrm} [1] {\mathrm{#1}}
\nc {\half} {\mbox{$\frac{1}{2}$}}
\nc {\thal} {\mbox{$\frac{3}{2}$}}
\nc {\fial} {\mbox{$\frac{5}{2}$}}
\nc {\la} {\mbox{$\langle$}}
\nc {\ra} {\mbox{$\rangle$}}
\nc {\etal} {\emph{et al.}}
\nc {\eq} [1] {(\ref{#1})}
\nc {\Eq} [1] {Eq.~(\ref{#1})}
\nc {\Refc} [2] {Refs.~\cite[#1]{#2}}
\nc {\Sec} [1] {Sec.~\ref{#1}}
\nc {\chap} [1] {Chapter~\ref{#1}}
\nc {\anx} [1] {Appendix~\ref{#1}}
\nc {\tbl} [1] {Table~\ref{#1}}
\nc {\Fig} [1] {Fig.~\ref{#1}}
\nc {\ex} [1] {$^{#1}$}
\nc {\Sch} {Schr\"odinger }
\nc {\flim} [2] {\mathop{\longrightarrow}\limits_{{#1}\rightarrow{#2}}}
\nc {\textdegr}{$^{\circ}$}
\nc {\inred} [1]{\textcolor{red}{#1}}
\nc {\inblue} [1]{\textcolor{blue}{#1}}
\nc {\IR} [1]{\textcolor{red}{#1}}
\nc {\IB} [1]{\textcolor{blue}{#1}}
\nc{\pderiv}[2]{\cfrac{\partial #1}{\partial #2}}
\nc{\deriv}[2]{\cfrac{d#1}{d#2}}
\nc{\gsim}{\raisebox{-0.13cm}{~\shortstack{$>$ \\[-0.07cm]
      $\sim$}}~}
\nc {\bit} {\begin{itemize}}
	\nc {\eit} {\end{itemize}}

\title{$^{15}$C inelastic $1/2^+ \rightarrow 5/2^+$ excitation: A single-particle versus a collective process}
\author{C.~Beckman}
\affiliation{Facility for Rare Isotope Beams, Michigan State University, East Lansing, Michigan 48824, USA}
\affiliation{Department of Physics and Astronomy, Michigan State University, East Lansing, Michigan 48824, USA}
\author{M.~Catacora-Rios}
\affiliation{Facility for Rare Isotope Beams, Michigan State University, East Lansing, Michigan 48824, USA}
\affiliation{Department of Physics and Astronomy, Michigan State University, East Lansing, Michigan 48824, USA}\author{C.~Hebborn}
\affiliation{Université Paris-Saclay, CNRS/IN2P3, IJCLab, 91405 Orsay, France}
\affiliation{Facility for Rare Isotope Beams, Michigan State University, East Lansing, Michigan 48824, USA}
\affiliation{Department of Physics and Astronomy, Michigan State University, East Lansing, Michigan 48824, USA}
\author{F.~M.~Nunes}
\email{nunes@frib.msu.edu}
\affiliation{Facility for Rare Isotope Beams, Michigan State University, East Lansing, Michigan 48824, USA}
\affiliation{Department of Physics and Astronomy, Michigan State University, East Lansing, Michigan 48824, USA}

\date{\today}

\begin{abstract}
\begin{description}
\item[Background] The excitation of one-neutron halo nucleus $^{15}$C from the $1/2^+$ ground state to the $5/2^+$ first excited state was measured at Argonne National Laboratory by impinging $^{15}$C on a deuterated target at \mbox{$7.1 A$ MeV}. This data was then analyzed in the Distorted Wave Born Approximation using a rigid rotor model for the excitation.
\item[Purpose] Being a one-neutron halo, we expect a single-particle excitation to better represent the excitation of $^{15}$C rather than a collective process. We expect the breakup of $^{15}$C to influence the reaction mechanisms because of the low one-neutron separation threshold, 
which is close in energy to $^{15}$C's first excited state. 
The goal of this work is to explore various the reaction mechanisms to reinterpret the data of Ref.~\cite{exp-paper} for the inelastic excitation of $^{15}$C.
\item[Method] We solve the scattering problem assuming a three-body model $^{14}$C$+n+d$. We use the Continuum Discretized Coupled Channel method (CDCC) and compare the results with those obtained assuming 1-step DWBA with quadrupole deformation, as done in the original experimental analysis. We also use Bayesian uncertainty quantification to estimate the uncertainties in our predictions coming from the $n$-$d$ interaction.
\item[Results] We analyze both the elastic and inelastic angular distributions for $^{15}$C(d,d')$^{15}\text{C}^*$ at $7.1 A$ MeV. Our results show that $^{15}$C breakup effects are important. 
\item[Conclusions] While CDCC predicts the  elastic angular distribution correctly, it is not able to fully describe the experimental inelastic angular distribution. We discuss additional effects that may be responsible for the remaining discrepancy.
\end{description}
\end{abstract}

\maketitle

\section{Introduction}
\label{sec:intro}
Understanding how protons and neutrons arrange themselves inside a nucleus, and how this organization evolves for nuclei exhibiting an extreme ratio of neutrons to protons,  is  one of the overarching goals in nuclear physics~\cite{mottelson1999,VerneyReview}. A stringent test of our nuclear structure models are the transition strengths between two  states,  since they are  sensitive  to  the distributions of neutrons and protons inside both states.
For unstable systems, located away from the valley of stability, these transition strengths can be studied through reaction experiments in inverse kinematics, in which the nucleus to be studied is sent onto a target. Inelastic scattering reactions, in which the projectile and/or the target nuclei get excited during the reactions, are of particular interest. When performed on heavy targets, inelastic  reactions are driven by electromagnetic transition probabilities, while on light targets, the inelastic excitation is caused by both the nuclear and Coulomb interactions. Either way, to infer reliable information from inelastic scattering observables, one needs an accurate reaction model.

When the reaction is dominated by the Coulomb interaction, i.e., at low beam energy and forward angles, the electromagnetic transitions are accurately inferred using exact electromagnetic operators and a two-body description of the reaction relying on semi-classical trajectories~\cite{ReviewCOULEX1,ReviewCOULEX2}.  Reactions driven by the nuclear interaction are more challenging because semiclassical trajectories do not provide an accurate description of the reaction dynamics, requiring a full quantum mechanical description. Typically, one relies on simplified structure models to describe the projectile nucleus' excitation, using either a collective (e.g. \cite{keeley1998}) or a single-particle picture (e.g. \cite{summers2004}). In collective models, the excitation of the projectile is assumed to be driven by changes in the wavefunctions of several of its nucleons, i.e., acting collectively.  This is modeled using deformed, i.e., non-spherical, nuclear densities, characterized by deformation lengths, that can be directly related to inelastic transition strengths~\cite{fs_book}.  In single-particle models, one represents the projectile's excitation as a modification in the wavefunction of only one of its nucleons. Single-particle models therefore rely on a two-body description of the projectile, and a three-body description of the reaction.
The choice of the structural model to describe the projectile excitation is often driven by prior knowledge, e.g., structure predictions, and comparison with experimental data.

Recently,  elastic and inelastic scattering cross sections to the $5/2^+$ state for a $^{15}$C nucleus on a deuterated target were measured at Argonne National Laboratory (ANL)~\cite{exp-paper}. The joint analysis of both datasets, assuming a collective model, allowed the extraction of a quadrupole deformation parameter, obtained by scaling the theoretical predictions to the experimental data~\cite{exp-paper}. The shape of the elastic angular distribution was reasonably reproduced, but some discrepancies were observed in the inelastic scattering cross sections.  The choice of the theoretical model is somewhat counterintuitive, as $^{15}$C  is a one-neutron halo nucleus,   exhibiting an extremely clustered structure, in which one loosely-bound neutron decouples from a $^{14}$C core \cite{nunes2003}. Because of this strong clustering, the  excitation of one-neutron halo nuclei has often been described as a single-particle process involving only the valence neutron (e.g., Ref.~\cite{PhysRevLett.118.152502} for the inelastic scattering of $^{11}$Be). Moreover, previous works also emphasize the importance of couplings to the continuum of the valence neutron when analyzing reactions involving one-neutron halo nuclei~\cite{ReviewCDCC,Nick15C,PhysRevC.78.011601}.
Motivated by the discrepancy observed in Ref.~\cite{exp-paper} and the success of previous analysis of reactions using a single-particle description of $^{15}$C~\cite{Nick15C,PhysRevC.78.011601}, we revisit in this study the analysis of the data~\cite{exp-paper}. To this end, we use a single-particle model for the excitation of $^{15}$C and a three-body description of the reaction,  properly treating couplings to the $^{14}$C-$n$ continuum.

This paper is organized as follows. In Sec.~\ref{sec:cdcc}, we present the three-body reaction model considered and the Continuum Discretized Coupled Channel (CDCC) method~\cite{austern1987} used in this work.  We then   provide  numerical details and  the nuclear interactions used for this study in Sec.~\ref{sec:NumericalDetails}. We also discuss how we quantify the uncertainties of these input interactions  in Sec.~\ref{sec:mcmc}. We then compare in Sec.~\ref{sec:results} predictions obtained with  the single-particle  and   the collective models of $^{15}$C to the experimental data~\cite{exp-paper}. Because both  descriptions fall short when compared to the dataset of Ref.~\cite{exp-paper}, the effects of other reaction mechanisms not included in the model on the elastic and inelastic scattering cross-sections are discussed in Sec.~\ref{sec:discussion}. Finally, Sec.~\ref{sec:conclusions} contains the conclusions of this work.

\section{Theory and inputs}\label{sec:CDCC Method}

\subsection{The Continuum Discretized Coupled Channels Method}
\label{sec:cdcc}

\begin{figure}[H]
    \centering
    \includegraphics[width=0.30\textwidth]{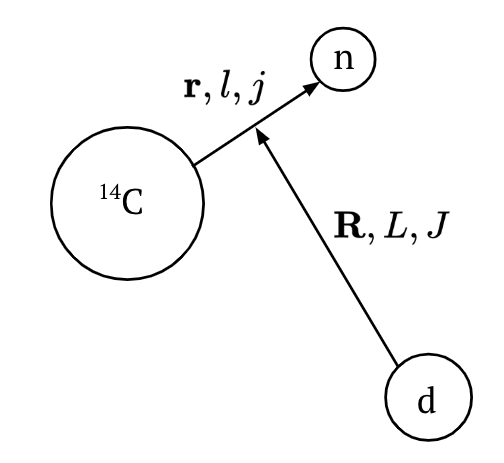}
    \caption{The 3-body $^{14}\text{C}+n+d$ coordinate system. 
    The relative orbital angular momentum of the $^{14}$C-$n$ and $^{15}\text{C}-d$ systems are $l$ and $L$, respectively, with $j$ and $J$ representing the corresponding total angular momenta.}
    \label{fig:coordinates}
\end{figure}

To model the excitation of $^{15}$C from $1/2^+$ to $5/2^+$ as a single-particle process (with the promotion of the valence neutron), it is necessary to include the internal degrees of freedom of the valence neutron explicitly: $^{15}$C$=^{14}\text{C}+n$. Thus, our reaction model for $^{15}$C$+d$ is based on the three-body Hamiltonian:
\begin{equation}
        H_{3b}=\hat{T}_{\ve R}+ H_{int}(\ve r)+U_{nd}(\ve{r}, \ve{R})+U_{dC}(\ve{r}, \ve{R}),
    \label{eq:h3b}
\end{equation}
where the internal Hamiltonian of the $^{15}$C projectile is defined as $H_{int}=\hat{T}_{\ve r}+V_{nC}(r)$. The interaction $V_{nC}$ is between the $^{14}$C core and the valence neutron, whereas $U_{nd}$ and $U_{dC}$ are the pairwise optical potentials between each fragment in the projectile, the $^{14}$C core and the valence neutron $n$,  and the deuteron target $d$. The Hamiltonian of Eqn.~\ref{eq:h3b} is expressed in terms of the Jacobi coordinates defined in Fig.~\ref{fig:coordinates}.

Due to the loosely-bound nature of $^{15}$C, as evidenced by the low neutron separation threshold of 1.218 MeV \cite{NNDC-15C}, we are interested in studying the effect of breakup on the inelastic process. Therefore,
we must include the possibility of $^{15}$C populating unbound states. 
For this purpose, the Continuum Discretized Coupled Channel method (CDCC) \cite{austern1987} is used. 
We note that the eigenstates of the internal Hamiltonian, defined by $ H_{int} \phi(\ve{r}) = \varepsilon \phi(\ve{r})$, form a complete set. Thus, we solve the three-body Schr\"odinger Equation $H_{3b} \Psi = E \Psi$ with scattering boundary conditions, by expanding the three-body incoming state in terms of these $\varphi(\ve{r})$,
\begin{equation}
        \Psi(\ve{r},\ve{R})=\sum_b \phi_b(\ve{r})\psi_b(\ve{R})+\int\phi_{\ve{k}}(\ve{r})\psi_{\ve{K}}^{\ve{k}}(\ve{R})d\ve{k} .
\label{eq:wfn3b}
\end{equation}
This expansion in Eqn.~\ref{eq:wfn3b} includes all bound states ($b$) as well as an integral over all the continuum states (represented by the $n-^{14}$C relative momentum $\ve k$). To make the problem tractable and square-integrable,  we use, instead of the scattering waves $\phi_{\ve{k}}(\ve{r})$, wave-packets representing each energy bin  $\tilde \phi_i(\ve{r})$, such that the three-body wavefunction becomes
\begin{equation}
        \Psi^{\rm CDCC}(\ve{r}, \ve{R})=\sum_{i,LJ}\tilde{\phi}_{i}(\ve{r})\psi_{iLJ}(\ve{R}) \; .
    \label{eq:wfncdcc}
\end{equation}
This sum over angular momentum $(LJ)$ and projectile excitation energy $(i)$ is truncated once convergence is reached on the desired observable (in our case, angular distributions for both elastic and inelastic cross sections). 
For more details on the CDCC method, see Ref.~\cite{fs_book}.

\subsection{Numerical Details}
\label{sec:NumericalDetails}

In Table \ref{tab:interactions}, we summarize the parameters used in the CDCC calculations. For the $^{14}$C-$d$ interaction, we use the optical potential developed in Ref.~\cite{ancai}. The $^{14}$C-$n$ interaction is adjusted to properties of $^{15}$C:
the real volume depth $V$ was adjusted to reproduce the neutron separation energy in the ground state ($S_n=1.218$ MeV) and the spin-orbit depth, $V_{so}$ was adjusted to reproduce the neutron separation energy in the first excited state ($S_n=0.478$ MeV) \cite{NNDC-15C}. With this potential, we obtain an asymptotic normalization constant (ANC) for the $^{15}$C ground-state of  $\mathcal{C}^2=1.48$ fm$^{-1}$, which is comparable to the experimentally determined value for the ANC in \cite{ANC_15C} (1.74$\pm$0.11 fm$^{-1}$).  
For the $n$-$d$ interaction we take a Yukawa form, inspired by Ref.~\cite{yukawa}. The parameters of this Yukawa interaction were fit on $p$-$d$ data \cite{pd_data} using the standard $\chi^2$ minimization procedure in {\sc sfresco} \cite{fresco}. We use $p$-$d$ data because the available $n$-$d$ data around $7.1$~MeV \cite{nd_data} does not constrain the parameters. Recognizing that this $n$-$d$ interaction carries large uncertainties, we  performed a full Bayesian calibration to estimate the magnitude of the uncertainties on the observables of interest (see Sec. \ref{sec:mcmc}).


\begin{table}[t!]
    \captionsetup{justification=raggedright, singlelinecheck=false}
\begin{tabular}{l | ccccccccc} 
\hline 
\hline \hline
 \rule{0pt}{3ex}	 &$V$ & $r_V$& $a_V$ &$V_{so}$ & $r_{so}$ &$a_{so}$ \\ 
  \rule{0pt}{3ex}	&[MeV]& [fm]&    [fm]&   [MeV] &     [fm]&        [fm]  \\ 
 \hline 
 \hline
 \rule{0pt}{3ex} $U_{nC}$ & 77.79 & 0.991 &  0.51 & 7.27 & 1.210 & 0.65 \\ \hline
 \rule{0pt}{3ex} $U_{nd}$ & 33.26 & 2.708 & $--$& $--$ & $--$ & $--$\\ \hline
 \rule{0pt}{3ex} $U_{dC}$ & 89.97 & 1.149 & 0.75 & $--$ & $--$ & $--$ \\ \hline 
  \hline 
  \rule{0pt}{3ex} & $W$ & $r_{w}$ & $a_{w}$ & $W_{d}$ & $r_{wd}$ & $a_{wd}$ \\ 
  \rule{0pt}{3ex}	&[MeV]& [fm]&    [fm]&   [MeV] &     [fm]&        [fm]  \\ \hline 
  \hline

  \rule{0pt}{3ex}$U_{dC}$ & 1.99 & 1.346 & 0.57 & 10.3 & 1.40 & 0.68 \\
\hline
\hline \hline
\end{tabular}	
\caption{Parameters for the interactions used in our calculation: the $^{14}$C-$d$ (core-target) optical potential is of Woods-Saxon form with finite-range Coulomb with $r_C=1.3$ fm, the $^{14}$C-$n$ bound state is also of Woods-Saxon form and contains a spin-orbit interaction, the $n$-$d$ interaction is of Yukawa form. Radii are obtained from scaling with the mass: $R_x=r_x\, 14^{1/3}$ for $U_{dC}$ and $U_{nC}$. No additional mass scaling is used for the range of the $U_{nd}$ Yukuwa interaction.}	
\label{tab:interactions}
\end{table}

CDCC calculations were performed with the code {\sc fresco} \cite{fresco}. The converged model space includes total angular momentum $J$ ($^{15}$C-$d$) up to $J_{max}=40$, a radial grid step size of $R_{step}=0.1$ and a matching radius of $R_{match}=100$ fm. We include up to $Q=4$ in the multipole expansion of the CDCC coupling potentials. 
The continuum was discretized in energy bins of 0.5 MeV up to a maximum energy of 10 MeV. For convergence of the elastic and inelastic cross-sections, $d\sigma/d\Omega$, we include all $^{14}$C$-n$ partial waves up to  $l_{max}=5$. For simplicity, in our calculations, we assume the spectroscopic factor for both bound states in $^{15}$C is unity.


\subsection{Quantifying uncertainties of  $U_{nd}$}
\label{sec:mcmc}


Through various tests, we identified that the $n$-$d$ interaction is the most uncertain in the reaction model considered here. For this reason, we chose to perform a Bayesian analysis to quantify its parametric uncertainties and propagate them to the $^{15}$C reactions. We use the same methods as in Refs.~\cite{lovell_bayes} and \cite{ManuelANC}. 
The prior is taken to be a Gaussian distribution with a 20\% width, centered on the central values of the parameters $V_{0}$ and $r_{0}$ of the $n$-$d$ Yukawa potential from the original fit using {\sc sfresco} (values shown in Table \ref{tab:interactions}).   The posterior distribution is sampled using Markov-Chain Monte Carlo (MCMC). At each step, the proposed parameter set is accepted or rejected according to,

\begin{equation}
    \begin{aligned}
        \frac{p(H_{f})\mathcal{L}(D|H_{f})}{p(H_{i})\mathcal{L}(D|H_{i})} > y \quad 
        y \in [0,1] \;.
    \end{aligned}
    \label{likelihood criteria}
\end{equation}
Here, the subscript $i$ denotes a given step in the MCMC chain; the probability of the prior distribution at this step $i$ is $p(H_{i})$ with likelihood $\mathcal{L}(D|H_{i})$; and the new set of parameters, denoted by subscript $f$, have prior probability $p(H_{f})$ and likelihood $\mathcal{L}(D|H_{f})$. 

The new set of parameters is selected from the normal distribution defined by $x_{f}\propto \mathcal{N}(x_{i},\epsilon x_{0})$, where $\epsilon$ is the MCMC step-size hyperparameter (here $\epsilon = 0.007$) and $x_0$ is the prior center. We use a standard likelihood $e^{-\chi^2/2}$, where the $\chi^{2}$ is defined as:
\begin{equation}
    \begin{aligned}
        \chi^{2}=\sum_{i=0}^{N}\frac{(\frac{d\sigma}{d\Omega}^{th}_i-\frac{d\sigma}{d\Omega}^{exp}_i)^{2}}{(\Delta\sigma^{exp}_i)^{2}+(M\frac{d\sigma}{d\Omega}^{exp}_i)^2}.
    \end{aligned}
    \label{chi squared}
\end{equation}
Here, $\Delta\sigma^{exp}$ is the experimental error, and $\sigma^{th}$ and $\sigma^{exp}$ are the predictions of the theoretical model and the experimental values, respectively. Note that $\Delta\sigma^{exp}$ contains both statistical and systematic reported errors.  $M$ is a hyperparameter introduced to account for unaccounted for uncertainties, such as model error. We set it to $M=0.35$, that provides a good empirical coverage \cite{pruitt2024}. 
For the calibration of the $U_{nd}$ potential the model parameters are $V_{v}$ and $r_{v}$. The experimental data used for the calibration is the dimensionless ratio to Rutherford elastic angular distribution of the $d(p,p)d$ reaction \cite{pd_data}. 

\begin{figure}[h!]
    \centering
    \captionsetup{
  justification=justified,
  singlelinecheck=false
}
    \includegraphics[width=0.9\linewidth]{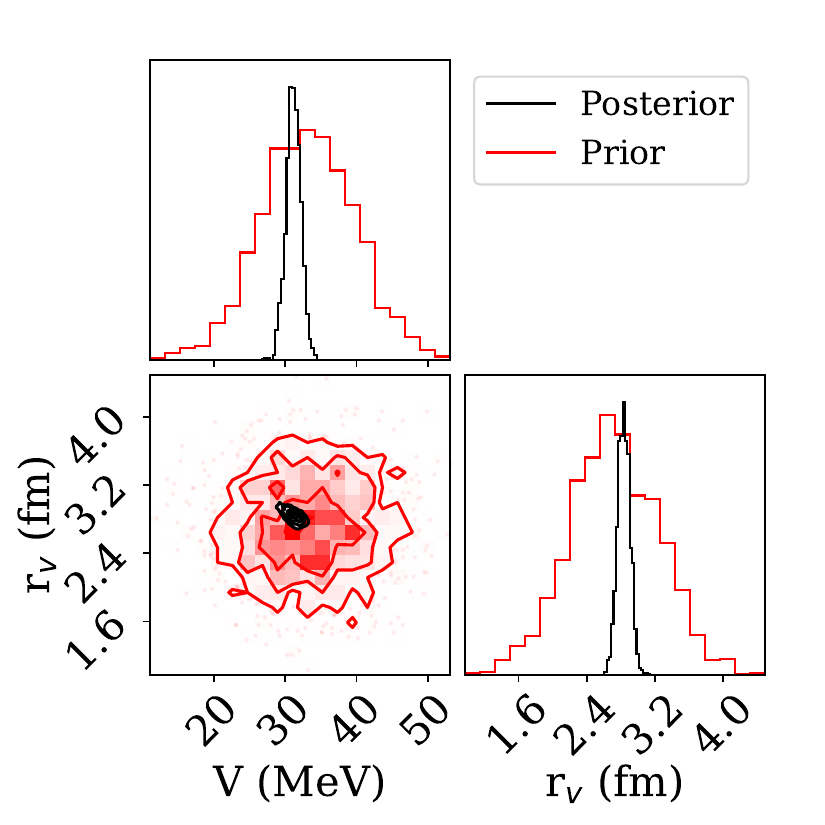}
    \caption{The corner plot for the calibration of $p-d$ data \cite{pd_data}: the prior distributions, centered at $V=33.26$ MeV and $r_{v}=2.708$ fm with a 20\% width, are in red and the resulting posterior distributions are in black.}
    \label{fig:cornerplots}
\end{figure}

For the MCMC sampling, we take a $500$ samples for burn-in and accept $1600$ parameters that satisfy condition \ref{likelihood criteria}. Fig. \ref{fig:cornerplots} shows the corner plot for the parameters that were calibrated in the procedure: along the diagonal we show the posterior distribution for the depths $V$ and radius $r_v$ of the $n-d$ interaction (in black), and compare it with the prior distribution (in red) included in the MCMC calculation. These are histograms representing the probability for each parameter, the peak is the most likely value and the width reflects the parameter uncertainty. Because the posterior is much narrower than the prior, we conclude the data used is able to significantly constrain these two parameters. The sampled region is shown in the off diagonal panels; here again the black area is significantly reduced compared to the red area, indicating that the data does reduce the phase space considered. Also, the fact that the black area is somewhat elongated and not circular indicates that these two parameters are negatively  correlated.


\section{Results}
\label{sec:results}

\begin{figure}[t!]
    \centering

    \includegraphics[width=0.45\textwidth]{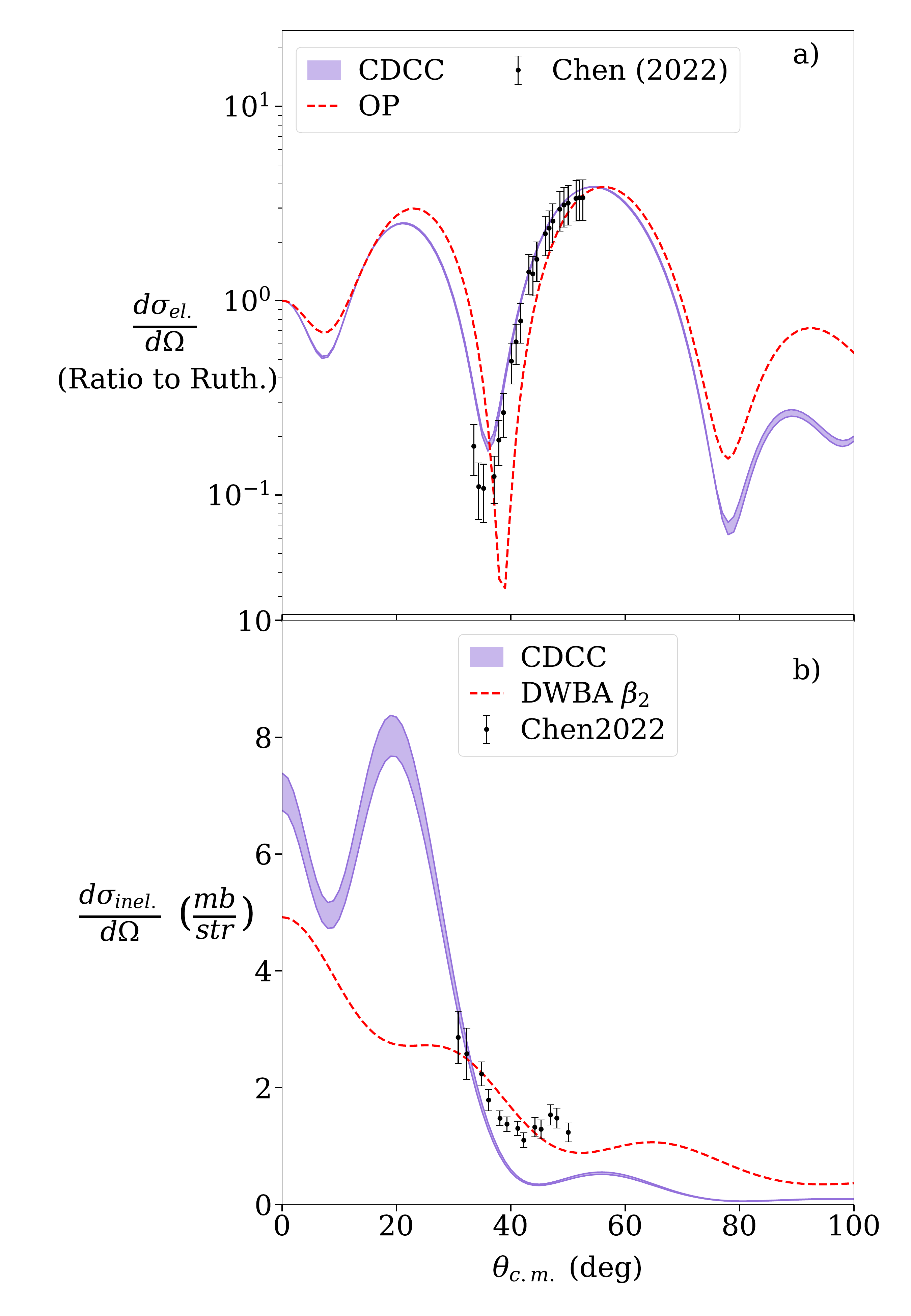}
    
    \caption{Calculations of the angular distributions for $^{15}$C on deuteron at $7.1A$ MeV elastic scattering (a) and inelastic scattering (b). The full CDCC (purple bands) are compared to: a) an optical models calculation and b) a DWBA calculation assuming a collective model with $\beta_{2}=0.35$  (red dashed lines). In black are the $d(^{15}$C, $^{15}$C$^{*})d'$  data of Ref.~\cite{exp-paper}, with error bars showing the sum of systematic and statistical error. }
    \label{fig:cdcc-def}
\end{figure}

As aforementioned, the main goal of this work is to develop a single-particle excitation model that includes breakup states of $^{15}C$, and compare the results with those using the collective model of $^{15}C$ in Ref.~\cite{exp-paper}. Fig. \ref{fig:cdcc-def}a) shows the elastic angular distribution for $^{15}$C on a deuteron target at $7.1A$~MeV as a function of scattering angle in the center of mass frame(ratio to Rutherford). The purple band is the $68$ \% credible interval for the CDCC calculation (the uncertainties pertain to the $n$-$d$ interaction only). This three-body model calculation is compared to a two-body optical-potential model, which is represented by the dashed-red line (labeled OP in Fig.\ref{fig:cdcc-def}a)). The red dashed line assumes the deuteron optical potential from the widely used global parametrization of Ref.~\cite{ancai}. The largest differences between these two approaches are seen at backward angles, where no data has been measured. For the measured angular region, the CDCC diffraction pattern shows a shift to smaller scattering angles compared to the single-channel approach used in Ref.~\cite{exp-paper}. The predictions from CDCC show a better agreement with the elastic data than the results using a deuteron optical model, including the reproductions of the pronounced minimum around $35 ^\circ$. This result is not surprising: as a halo nucleus, we expect $^{15}$C excitations, to the $5/2^+$ and the continuum, to modify the angular distribution. The optical model of \cite{ancai} was developed for stable nuclei and it is not expected to work well on exotic nuclei.

In Fig. \ref{fig:cdcc-def}b), we present the calculations for the inelastic channel populating the $5/2^+$ state in $^{15}$C. Again, the purple band corresponds to the $68$\% credible interval for the CDCC predictions (again, the uncertainties are from the $n$-$d$ interaction alone), and the dashed-red line corresponds to a 1-step DWBA calculation assuming 
a collective model for the $1/2^+ \rightarrow 5/2^+$ excitation, where the quadrupole deformation length is fitted directly to the inelastic data (equivalent to the analysis of Ref.~\cite{exp-paper}). The DWBA results shown in Fig. \ref{fig:cdcc-def}b) have a deformation parameter $\beta_2=0.35$. 
Both CDCC and  DWBA-$\beta_2$ results agree with the data at the lowest measured angles. However,  DWBA-$\beta_2$ calculation was specifically fitted to reproduce the magnitude of the cross section at these angles. In contrast, CDCC is a true prediction; it was not fit to either elastic or inelastic data shown in Fig.\ref{fig:cdcc-def}). The most important point to make though, is that neither calculation reproduces the angular behavior in the data for the largest measured angles. 
The CDCC prediction does get the correct angle for the second minimum, slightly above $40^{\circ}$, but it does not produce a third peak consistent with the data.

\begin{figure}[h!]
    \centering
    \captionsetup{justification=raggedright, singlelinecheck=false}
    \includegraphics[width=0.45\textwidth]{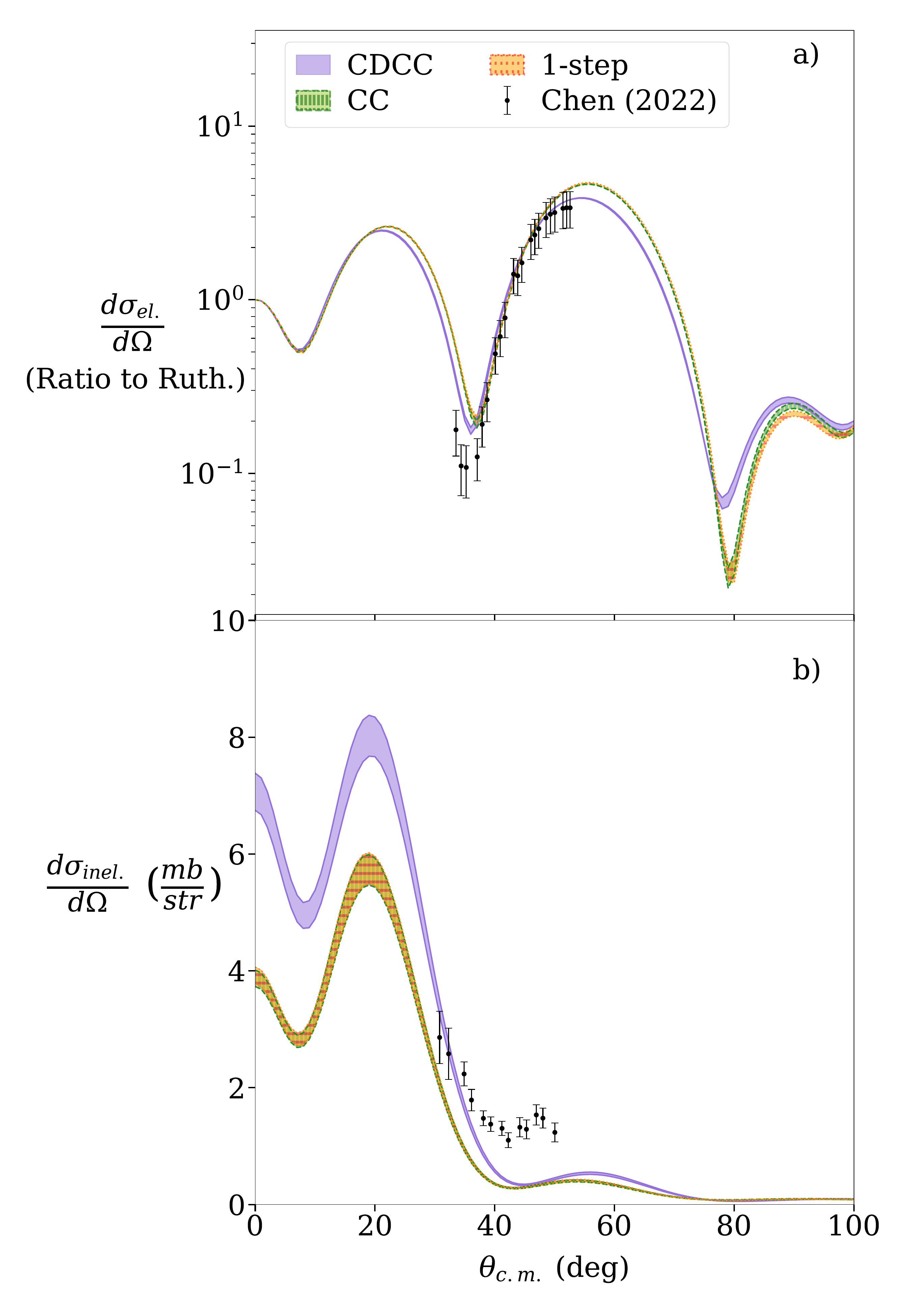}
    \caption{Different reaction mechanisms for reactions of $^{15}$C on deuterons at $7.1A$ MeV, elastic scattering (a) and inelastic scattering (b):  CDCC (purple bands) are compared to the coupled channel calculations including only bound states (CC, green bands) and with single-step DWBA (orange bands). }
    \label{fig:mechanisms}
\end{figure}

The shortcomings in the reaction predictions shown in Fig.\ref{fig:cdcc-def}b)
lead us to further inspect the mechanisms involved in this reaction. We next investigate the effects of the various couplings in the original CDCC calculations.
In a full CDCC calculation, the neutron can bounce around the various states in the model space before ending up in the exit channel (elastic, inelastic, or breakup). But now we also consider two simplified reaction mechanisms: first, we assume the single-particle excitation happens in one-single step; it corresponds to 1-step DWBA, with single-particle couplings (labelled 1-step), and second, we allow for the neutron to go back and forth between the ground state and the $5/2^+$ but not the continuum; a coupled channel calculation including bound states only (labelled  CC). The predictions for the elastic and inelastic cross sections for these two reaction mechanisms are compared to the full CDCC in Fig. \ref{fig:mechanisms}. 
By comparing the CC distribution (green) with the 1-step result (orange), we immediately conclude that higher-order effects between the two bound states ($1/2^+ \leftrightarrow 5/2^+$) have a negligible effect on the elastic and inelastic cross sections. However, the difference between the CDCC and the CC $68$\% credible intervals (purple versus green bands) are significant, indicating the importance of breakup states. The breakup channels shift the elastic angular distribution slightly toward smaller angles and the inelastic angular distribution toward larger angles. Without the breakup couplings, the agreement with the inelastic experimental data would have been worse.

We have further investigated whether other details in our three-body model could explain the disagreement between the inelastic CDCC prediction between $40^{\circ}-50^{\circ}$ and the data. The $^{14}$C-$d$ interaction does reproduce elastic data and should carry much smaller uncertainty than the $n$-$d$. The $^{14}$C-$n$ interaction has been well constrained for $l=0$ and $l=2$ but not in the $l=1$ channels. For this reason, we repeated the full CDCC calculations assuming different depths for the $p$-wave: for one case we take the  $p$-wave strength to reproduce the $^{13}$C+$n$ separation energy, and in another case we assume the  $p$-wave strength to be zero. While these changes produce large effects on the breakup cross sections, the effect is hardly noticeable on the elastic and the inelastic angular distributions.
Therefore, the uncertainty on the $^{14}$C-$n$ interaction in the $p$-wave channel cannot explain the disagreement with the inelastic data.

Finally, although we use the separation energy of the bound states to constrain the $^{14}$C-$n$ interaction, there is still a remaining ambiguity in the choice of radius and difuseness. We have repeated the CDCC calculations assuming a couple of different $(V,V_{so},r_v,a_v)$ sets, while still imposing the separation energies of $^{15}$C bound states, pulling from the posterior distributions obtained in \cite{ManuelANC}, which had a constraint of the $^{15}$C ground state ANC. While outside the range of the data, these changes can produce significant variation in the elastic or inelastic cross sections (at backward angles and forward angles respectively), the changes on the cross section are minor in the angular range $40-50$ degrees, and therefore the disagreement discussed above with the inelastic data remains.

We note that the inelastic cross section scales mostly with the product of the ANCs for the ground state and the first excited state (the inelastic reaction at these energies is largely peripheral), and therefore constraining these two quantities independently will further reduce the uncertainties on the $^{14}$C-$n$ interaction.

The persistent disagreement with the inelastic data seems to suggest that there are other reaction mechanisms relevant to the process that have not been included in our model space. This is what we discuss next.

\section{Discussion}
\label{sec:discussion}

In this section, we explore other reaction mechanisms, beyond those discussed in Sec. \ref{sec:results}. These include deuteron breakup, and transfer  $(d,t)$  and  $(d,p)$.

\begin{figure}
    \centering
    \captionsetup{justification=raggedright, singlelinecheck=false}
    \includegraphics[width=\linewidth]{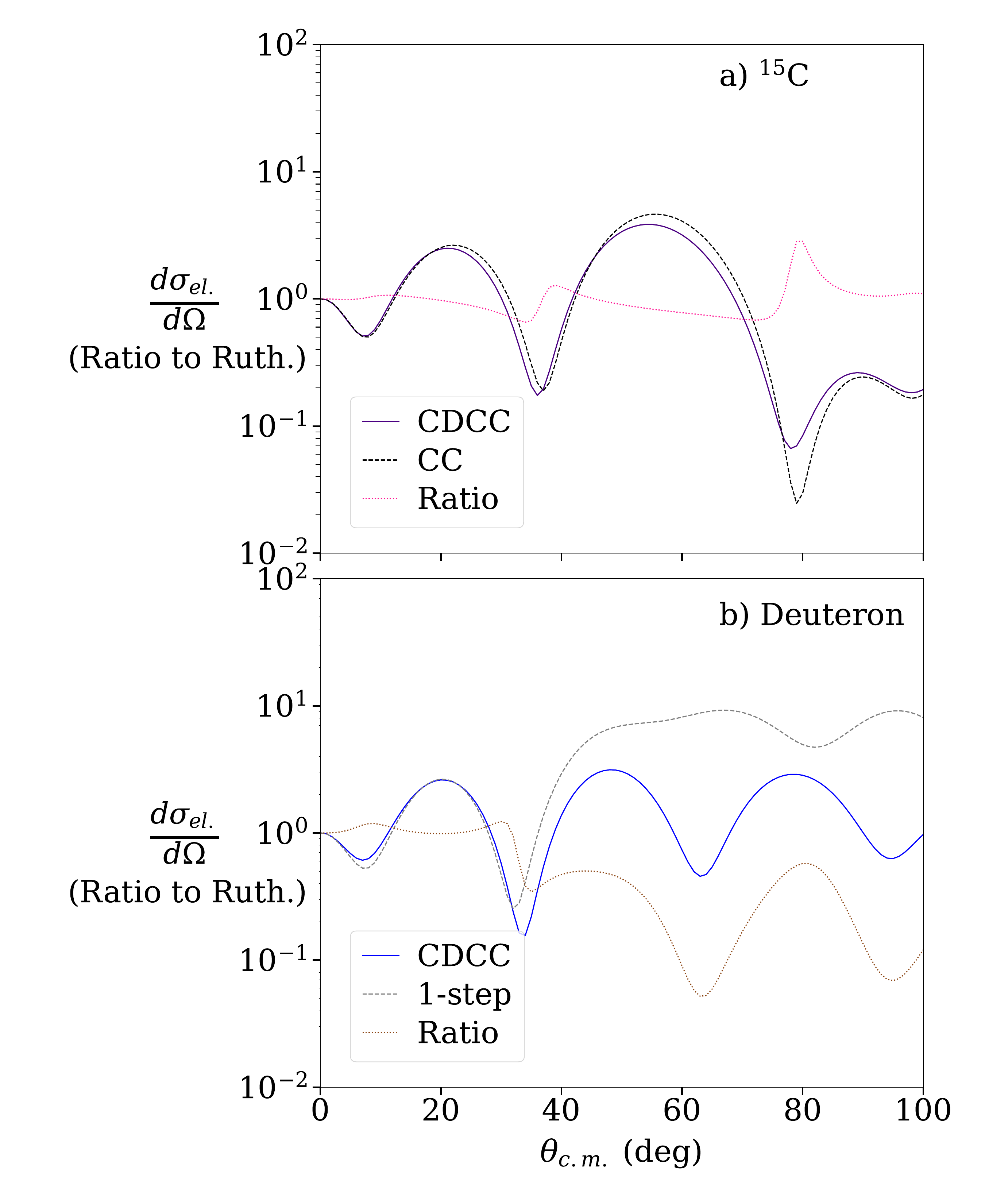}
    \caption{Assessment of the importance of breakup channels on the $^{15}$C$+d$ elastic angular distribution: a) full CDCC calculations (solid purple) and coupled-channel  including only the two bound states (dashed black) for $^{15}$C breakup; b) full CDCC calculations (solid blue) and 1-step (dashed gray) $d$ breakup.  The corresponding ratios are also plotted.}
    \label{fig:RatioBreakupComp}
\end{figure}

The deuteron is itself loosely bound; deuteron breakup may affect the inelastic reaction. The ideal framework, a four-body model composed of $^{14}$C$+n+n+p$, is beyond the scope of this work. Thus, to assess the significance of deuteron breakup, and compare it with that of $^{15}$C breakup, we perform CDCC calculations for $^{15}$C$(d,np)^{15}$C at $7.1A$ MeV. The interactions used in the CDCC calculations are: $V_{pC}$ and $V_{nC}$ from Ref.~\cite{koning_delaroche}, and $V_{np}$ from Ref.~\cite{gaussian_np}. We repeat those calculations without continuum couplings and take the ratio of the corresponding elastic distributions. We compared this ratio with that obtained when considering $^{15}$C breakup in Fig.~\ref{fig:RatioBreakupComp}: the top panel (a) concerns $^{15}$C breakup while the bottom panel (b) concerns deuteron breakup. In the range where this reaction was measured, deuteron breakup has a larger effect on the elastic than $^{15}$C breakup. Considering specifically the range where CDCC does not match the inelastic data, the effect of deuteron breakup on the elastic is of similar magnitude to that of $^{15}$C breakup. We note that there could be a complex interplay of the deuteron continuum and the $^{15}$C continuum, making the analysis of the inelastic experiment less straightforward. Such effects have been discussed for example in the context of transfer  \cite{moro2009} and breakup reactions \cite{Descouvemont2016}.

In addition, to explore the possibility that the inelastic process couples to transfer channels, we consider the magnitude to $^{15}$C$(d,p)$ to various states in $^{16}$C and $^{15}$C(d,t)$^{14}$C(g.s.) cross sections, as data for these are available in Ref.~\cite{dp_transfer} and Ref.~\cite{dt_transfer_kay_paper} respectively.
The angular distributions for $(d,p)$  are larger than the inelastic cross sections, although they are only measured for $\theta_{CM} < 25 ^{\circ}$. 
The $(d,t)$  cross sections have similar magnitude to the inelastic channel $^{15}$C$(d,d')$. 
This suggests that complex reaction mechanisms, coupling the elastic/inelastic channels with transfer channels, may be taking place.

Finally, the CDCC model considered in this work assumed the $^{14}$C is inert. The first excited states of $^{14}$C are just above 6 MeV, so these channels are indeed open for the $^{15}$C$(d,d')$ reaction at $7.1A$ MeV. It may thus be useful to explore the effect of core excitation in an XCDCC framework \cite{xcdcc}. This is beyond the scope of this work.

\noindent

\section{Conclusions}
\label{sec:conclusions}

In this work, we consider the inelastic excitation of a halo nucleus when reacting with a light target, and contrast a model that assumes the process happens as a  single-particle excitation with a model that assumes a collective excitation. Specifically, we consider the reaction of $^{15}$C on deuterons at 7.1$A$ MeV as measured by the HELIOS setup at ANL~\cite{exp-paper}. We use the CDCC three-body model $^{14}$C$+n+d$ and include, in addition to the two bound states, full couplings to the continuum of $^{15}$C. The pairwise interactions were fitted to properties of the subsystems but, 
due to the ambiguity involved in extracting a $n$-$d$ interaction, a Bayesian calibration was done to estimate the uncertainties in our predictions.

First, our results show that inclusion of $^{15}$C breakup effects are important in the $^{15}$C+$d$ reactions considered: the inclusion of $^{15}$C breakup shifts the elastic and inelastic angular distribution, bringing them in better alignment with the data. The predictions from CDCC are significantly different than those obtained in DWBA using a quadrupole collective coupling for the excitation. 
While the CDCC single-particle predictions describe better the elastic scattering than the collective approach, and agree well with the measured angular distribution, the situation for the inelastic channel is mixed. The angular dependence of the inelastic cross section predicted by CDCC does not agree with the data, even considering the uncertainty resulting from the $n$-$d$ interaction. Even though the magnitude of the  DWBA cross section predicted with the collective model is adjusted to fit the data, its angular dependence does not follow the details of the diffraction pattern measured either.

In our study, we inspect several other reaction mechanisms that could justify the disagreement seen. We found that the inelastic reaction under study could be affected by deuteron breakup and by coupling to transfer channels, since the cross sections for these processes are of similar magnitude to the inelastic data. In order to test the hypothesis of single-particle versus collection excitation of a halo nucleus, replacing the deuteron target by a proton target removes much of the complexity discussed and offers a more straightforward analysis.

\begin{acknowledgements} We thank Ben Kay, Daniel Bazin and Yassid Ayyad for useful discussions regarding the experimental data. We thank the few-body reactions group for feedback on this project. We acknowledge the high performance computing resources at iCER (MSU). This project  received financial support from the CNRS through the AIQI-IN2P3 project and was supported in part by the U.S. Department of Energy grant DE-SC0021422 and the National Science Foundation CSSI program under award No. OAC-2004601 (BAND Collaboration).
\end{acknowledgements}

\bibliographystyle{apsrev}
\bibliography{Biblio}
\appendix

\end{document}